\documentclass[journal=jacsat,manuscript=article]{achemso}

\usepackage[version=3]{mhchem} 
\usepackage{graphicx, glossaries, ulem, xcolor, hyperref, multicol}
\usepackage[separate-uncertainty = true]{siunitx}

\newcommand{\vbg}{\ensuremath{V_\mathrm{BG}}}
\newcommand{\vtg}{\ensuremath{V_\mathrm{TG}}}
\newcommand{\vsd}{\ensuremath{V_\mathrm{SD}}}
\newcommand{\mstar}{\ensuremath{m^\mathrm{*}}}
\newcommand{\subtxt}[2]{\ensuremath{#1_\mathrm{#2}}}

\newacronym{dqd}{DQD}{double quantum dot}
\newacronym{qd}{QD}{quantum dot}
\newacronym{qw}{QW}{quantum well}
\newacronym{soi}{SOI}{spin--orbit interaction}
\newacronym{hh}{HH}{heavy-hole}
\newacronym{lh}{LH}{light-hole}
\newacronym{2dhg}{2DHG}{two dimensional hole gas}
\newacronym{haadf}{HAADF}{high-angle annular dark-field}

\newacronym{sige}{SiGe}{silicon-germanium}
\newacronym{ge}{Ge}{germanium}
\newacronym{si}{Si}{silicon}
\newacronym{pt}{Pt}{platinum}
\newacronym{al}{Al}{aluminium}

\newacronym{sem}{SEM}{scanning electron microscopy}
\newacronym{tem}{TEM}{scanning tunnelling microscopy}
\newacronym{cvd}{CVD}{chemical vapor deposition}
\newacronym{ald}{ALD}{atomic layer depostion}
\newacronym{rie}{RIE}{reactive ion etching}
\newacronym{sdh}{SdH}{Shubnikov–de Haas oscillations}
\newacronym{dhg}{2DHG}{two-imensional hole gas}
\newacronym{edx}{EDX}{energy dispersive X-Ray}
\newacronym{icp}{ICP}{inductively coupled plasma}

\DeclareSIUnit\torr{Torr}
\DeclareSIUnit\hour{h}

\author{Luigi Ruggiero}
\affiliation{Physics Department, University of Basel, Klingelbergstrasse 82, CH-4055 Basel, Switzeland}
\email{luigi.ruggiero@unibas.ch}
\author{Arianna Nigro}
\affiliation{Physics Department, University of Basel, Klingelbergstrasse 82, CH-4055 Basel, Switzeland}
\author{Ilaria Zardo}
\affiliation{Physics Department, University of Basel, Klingelbergstrasse 82, CH-4055 Basel, Switzeland}
\alsoaffiliation{Swiss Nanoscience Institute, Klingelbergstrasse 82, CH-4055 Basel, Switzeland}
\author{Andrea Hofmann}
\affiliation{Physics Department, University of Basel, Klingelbergstrasse 82, CH-4055 Basel, Switzeland}
\alsoaffiliation{Swiss Nanoscience Institute, Klingelbergstrasse 82, CH-4055 Basel, Switzeland}
\email{andrea.hofmann@unibas.ch}

\title[Backgate]
  {A backgate for enhanced tunability of holes in planar germanium}


\begin{document}


\begin{abstract}
    Planar semiconductor heterostructures offer versatile device designs and are promising candidates for scalable quantum computing. Notably,  heterostructures based on strained germanium have been extensively studied in recent years, with emphasis on their strong and tunable spin-orbit interaction, low effective mass, and high hole mobility. However, planar systems are still limited by the fact that the shape of the confinement potential is directly related to the density. In this work, we present the successful implementation of a backgate for a planar germanium heterostructure. The backgate, in combination with a topgate, enables independent control over the density and the electric field, which determines important state properties such as the effective mass, the $g$-factor and the quantum lifetime. This unparalleled degree of control paves the way towards engineering qubit properties and facilitates the targetted tuning of bilayer quantum wells, which promise denser qubit packing.
\end{abstract}

Compressively strained Ge \glspl{qw} in SiGe heterostructures \cite{} have become well known for their \gls{hh} ground states which have interesting properties for applications in quantum computing \cite{scappucci_germanium_2020, jirovec_singlettriplet_2021, hendrickx_fast_2020}
and research on hybrid superconductor/semiconductor devices\cite{hendrickx_gatecontrolled_2018, aggarwal_enhancement_2021, vigneau_germanium_2019}. In undoped Ge/SiGe heterostructures, hole mobilities of several million \si{\centi\meter^2/\volt\second} have been reached \cite{lodari_lightly_2022, myronov_holes_2023}, larger than those achieved in Si. However, by controlling the charge density with a top barrier alone, the density and the location of the charges within the \gls{qw} are always closely related. The states are pushed towards the top SiGe barrier and the confinement strength grows with density. The close vicinity of the confined states to the interface increases the importance of interface scattering and any finite wavefunction overlap with the SiGe barrier may lead to stronger decoherence mechanisms in qubit devices. Furthermore, the confinement potential and the electric field affect the amount of \gls{hh}-\gls{lh} splitting \cite{winkler_spin_2003, terrazos_theory_2021} and influence important properties such as the effective mass \cite{lodari_light_2019}, the spin--orbit interaction \cite{stano_heavyhole_2024} and the $g$-factor \cite{ivchenko_electronic_1997, terrazos_theory_2021}. 

We introduce a backgate to planar Ge/SiGe heterostructures, aiming to independently tune the density and the location of the hole wavefunction within the \gls{qw}. Backgates are routinely used in 2D material systems for a separate control of the density and displacement field \cite{novoselov_electric_2004, oostinga_gateinduced_2008, mak_observation_2009, zhang_direct_2009} and have been successfully applied to heterostructures for the purpose of increased device flexibility or measuring double \gls{qw} devices \cite{eisenstein_independently_1990, weckwerth_epoxy_1996, rubel_fabrication_1998, berl_structured_2016}. Integrating a backgate into a heterostructure is more intricate as the substrate is thick. It requires precise control over the thinning process so as to bring the backgate in close vicinity of the \gls{qw} to achieve an appreciable field-effect. 


\begin{figure}[ht!]
        \centering
        \includegraphics[width = 1.0\textwidth]{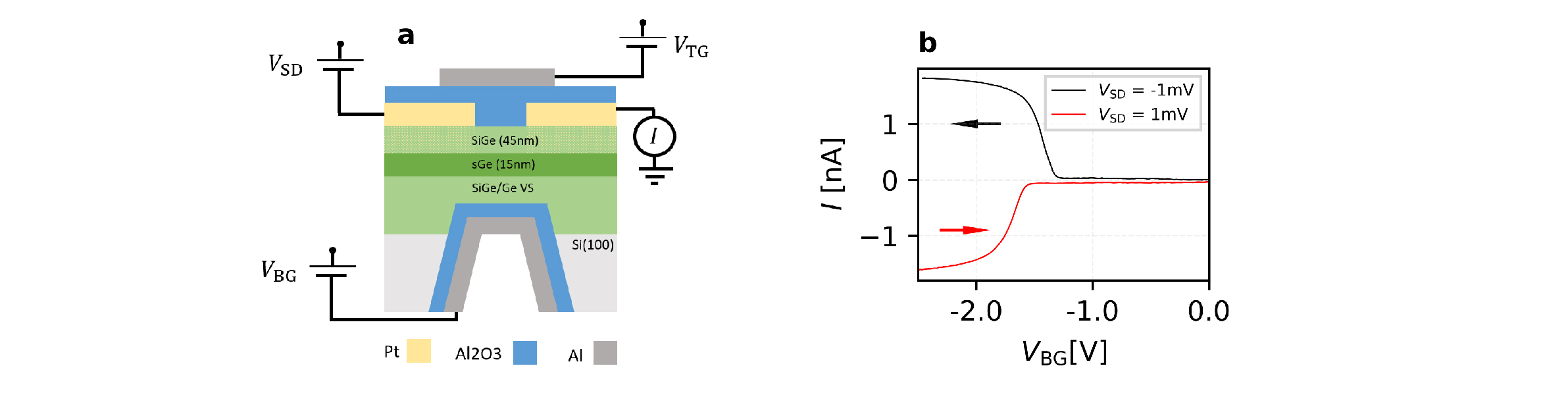}
        \caption{\textbf{Single gate device characterization.} (a) Sketch of the device and measurements setup. The flat region of the backgate is etched into the SiGe gradient buffer. (b) Accumulation curve of the current as a function of backgate voltage $\vbg$ performed at $\SI{4}{K}$ on a device with a backgate and without a topgate. The two curves are measured with opposite polarity of the applied source-drain voltage $\vsd$ and display sweeps in reverse direction.  }
        \label{fig:sample}
\end{figure}
Here, we make use of selective etching of the involved materials in order to place the backgate less than \SI{1}{\micro\meter} below the \gls{qw}, close enough to tune the \gls{2dhg} density with reasonable voltages. The heterostructure is schematically shown in Fig.~\ref{fig:sample}(a). Its bottom layers consist of a thick Si substrate, a Ge virtual substrate, and a reverse-graded buffer to reach a constant composition of \SI{80}{\percent} Ge in the barrier. The states of interest are the quasi-2D \gls{hh} states in the Ge \gls{qw}, which is confined from the top by another Si$_{0.2}$Ge$_{0.8}$ layer. The backgate is produced by wet etching the Si substrate with NaOH, an etchant that has a high selectivity to Si compared to Ge. The Al backgate is then fabricated on top of an insulating aluminium oxide layer. Ohmic contacts are fabricated from the top in order to measure the current induced in the \gls{2dhg} upon application of suitable gate voltages.

In a first step, we demonstrate the accumulation of a finite hole density within the Ge \gls{qw} solely due to a backgate, i.e. in absence of any topgate. Fig.~\ref{fig:sample}(b) shows the current $I$ flowing due to a bias voltage \subtxt{V}{SD} as a function of voltage \subtxt{V}{BG} applied to the backgate when the sample is cooled to \SI{4}{\kelvin}. As expected for an undoped heterostructure, no current is observed at zero backgate voltage. However, when sweeping \subtxt{V}{BG} negative, the current rises due to the accumulation of charges in the \gls{qw}. When sweeping \subtxt{V}{BG} in reverse direction, the depletion occurs at slightly more negative voltages. This observation has also been made for topgated devices \cite{massai_impact_2023}. In our case, the hysteresis is likely induced by charging of traps at imperfections in the bottom of the heterostructure or due to Ge segregation \cite{wan_highgecontent_2022}. While the hysteresis is not always reversible at low temperatures, the original state is retrieved after a thermal cycle to room temperature. The two curves are obtained with opposite sign of \subtxt{V}{SD} and the equal magnitude of measured current demonstrates the absence of sizeable leakage between the ohmic contacts and the backgate. In fact, we do not observe leakage down to \vbg = \SI{-7}{\volt}.

\begin{figure}[ht!]
        \centering
        \includegraphics[width = 1.0\textwidth]{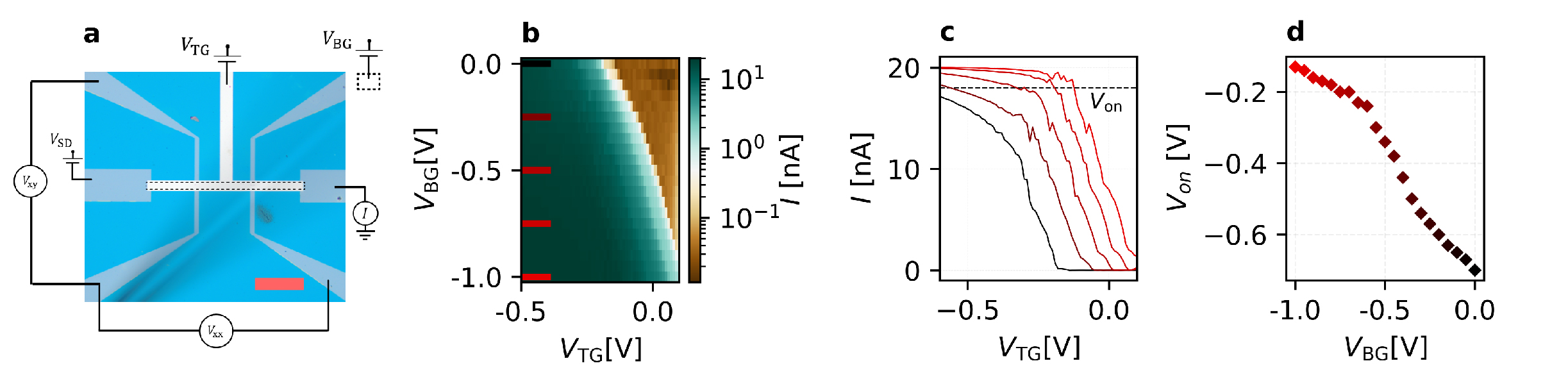}
        \caption{\textbf{Device details and accumulation curves with the top- and backgate.} (a) Confocal microscope picture of the hallbar device with a sketched measurement setup. The scale bar is \SI{100}{\micro\meter}. The diagonal line is a bond-wire. (b) 2D map of accumulation curves as function of $\vtg$ at different values of $\vbg$. (c) Traces at different $\vbg$ marked in (b). $V_{\mathrm{on}}$ is defined as $\vtg$ leading to \SI{90}{\percent} of the saturation current. (d) $V_{\mathrm{on}}$ as a function of $\vbg$.}
        \label{fig:backandtop}
\end{figure}
Having demonstrated the charge accumulation with a backgate alone, we move on to build devices with back- and topgates, with the aim to individually tune the density and the shape of the confinement potential. The optical micrograph in Fig.~\ref{fig:backandtop}(a) depicts a top-view of the sample, showing the topgate in white color, ohmic contacts in grey and indicating the location of the backgate with dashed lines. The design of the sample is limited by our fabrication process allowing only rectangular backgates and a precision of \SI{1}{\micro\meter} for the alignment between back and top. The backgate is slightly smaller than the topgate and the region of the \gls{2dhg} defined by the backgate is directly contacted with ohmic contacts. All of the following measurements are performed in a dilution refrigerator at base temperature of \SI{10}{\milli\kelvin} unless noted otherwise.

We first characterize the relative size of the field effect of the two gates by measuring the current through the device as a function of top- and backgate voltages, see Fig.~\ref{fig:backandtop}(b). The line-traces in Fig.~\ref{fig:backandtop}(c) show how the accumulation curves obtained with the topgate change with the value of applied backgate voltage. To allow a quantitative comparison, we define \subtxt{V}{on} as the topgate voltage where the current equals \SI{90}{\percent} of the saturation current. We compare in Fig.~\ref{fig:backandtop}(d) \subtxt{V}{on} for different values of \subtxt{V}{BG}. The slope amounts to \num{0.5}, suggesting that the backgate voltage has half the effect of the topgate voltage. This ratio is much bigger than our expectations based on the respective distances, \subtxt{d}{TG} and \subtxt{d}{BG}, of the gates to the \gls{qw} with a ratio of $\subtxt{d}{BG}/\subtxt{d}{TG} > 10$.

\begin{figure}[ht!]
        \centering
        \includegraphics[width = 1.0\textwidth]{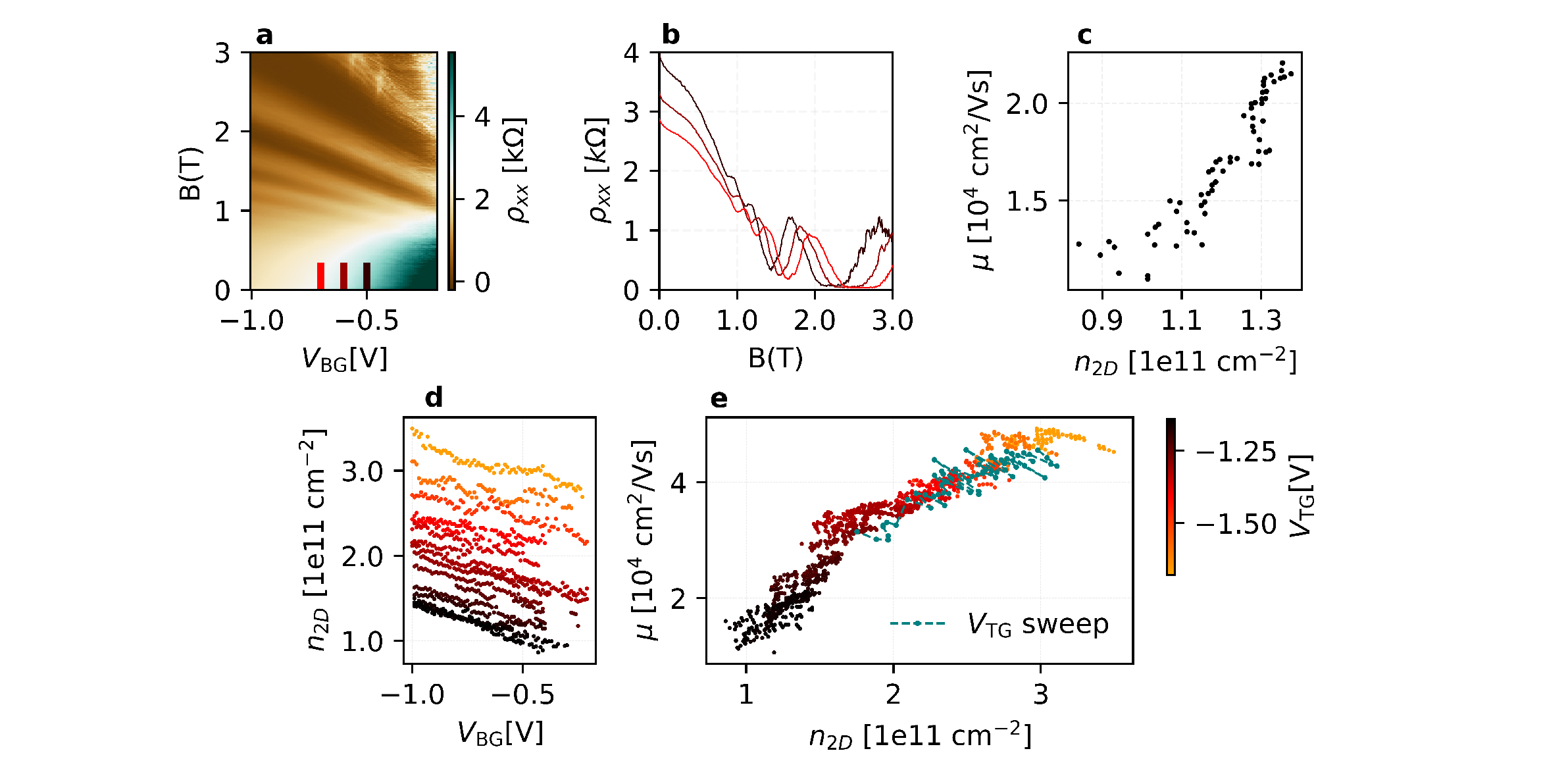}
        \caption{\textbf{Magnetotransport measurements with the $\vbg$} (a) \subtxt{\rho}{xx} as a function of $\vbg$ and magnetic field $B$. (b) Traces along different $\vbg$ as indicated in (a). (c) Mobility as a function of hole density. The density is extracted from \gls{sdh} oscillations in (a) and is then used for calculation of the mobility together with longitudinal resistance. (d) Densities as function of $\vbg$ at different $\vtg$. (e) Mobility as a function of density extracted from measurements at varying $\vtg$. The data extracted with sweeping $\vbg$ perfectly overlap with those obtained with $\vtg$ at $\vbg = \SI{0}{\volt}$.
}
        \label{fig:densitytune}
\end{figure}
In a different cooldown, we perform standard direct-current magnetotransport measurements to evaluate the efficiency of the backgate in tuning the \gls{2dhg}  density. A Landau fan is shown in \autoref{fig:densitytune}(a), which has been recorded by only sweeping the backgate voltage and keeping \subtxt{V}{TG} at a constant value just before accumulation. The line traces in (b) demonstrate the quantization into Landau levels when the longitudinal resistivity plateaus at zero. The density is extracted from the period of the \gls{sdh} at low fields and is then used to evaluate the mobility. As shown in \autoref{fig:densitytune}(c), the \gls{2dhg} density, \subtxt{n}{2D}, changes due to the voltage applied to the backgate. The mobility increases with density, which is typical in the low density regime, where particle scattering is still negligible compared to the effects of the potential homogeneity.

Aiming to further analyze the relative influence of the two gates onto the \gls{2dhg}, we perform magnetotransport measurements with increasingly negative values of \subtxt{V}{TG}. As shown in \autoref{fig:densitytune}(d), the density is linear in \subtxt{V}{BG}, demonstrating the field-effect induced by the backgate. Overall, \subtxt{n}{2D} increases with more negative \subtxt{V}{TG}, as more charges are accumulated by the topgate. 
From these traces, the relative field effect of the backgate compared to the topgate is estimated as $\sim\num{0.3}$. Meanwhile, the effect of the two gates on the mobility is very similar, as illustrated in \autoref{fig:densitytune}(e). There, we analyze the same magnetotransport data as for panel (d) to extract the mobility obtained when sweeping the backgate at constant \subtxt{V}{TG}. For a comparison, the green datapoints display the mobility extracted from an independent measurement where the backgate is kept at ground and magnetotransport measurements have been performed using only the topgate (see supplementary data). The two sets of data are in perfect agreement, indicating that applying voltages to the backgate does not deteriorate the \gls{2dhg}. 

\begin{figure}[ht!]
        \centering
        \includegraphics[width = 1.0\textwidth]{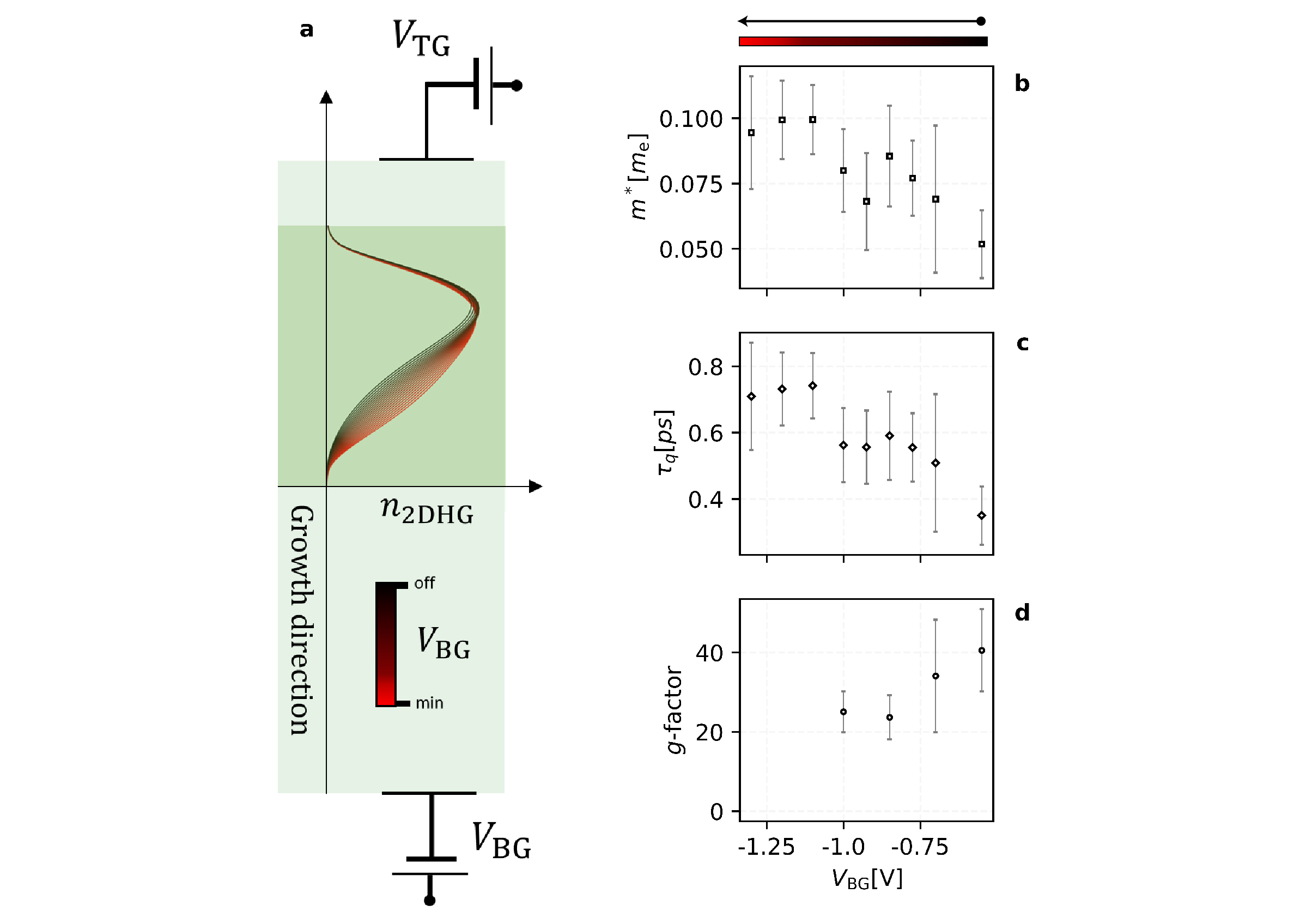}
        \caption{\textbf{Simulations and analysis of \gls{qw} properties.} (a) Finite element simulations of a simplified strained Ge/SiGe heterostructure. Density curves for different backgate voltages (from 0 to the most negative values) at fixed topgate voltage. (b) Effective mass $\mstar$, (c) quantum lifetime $ \subtxt{\tau}{q}$, (d) $g$-factor extraction for different values of $\vbg$ at constant hole density. 
}
        \label{fig:stateproperties}
\end{figure}
Using both gates, however, changes the confinement potential which in turn may affect properties such as the effective mass $\mstar$, the quantum lifetime \subtxt{\tau}{q} and the $g$-factor of the states under investigation. In fact, the topgate alone produces a triangular confinement potential with the wavefunction pushed towards the top barrier. Using both gates allows for a more symmetric configuration as shown in the \textit{nextnano} simulations of a simplified heterostructure in \autoref{fig:stateproperties}(a).

We follow Ref.~\cite{lei_electronic_2020} to extract the effective mass $\mstar$ from the temperature-dependence of the \gls{sdh} oscillations (see supplementary data) for a set configuration of top- and backgate voltages. Each data point in \autoref{fig:stateproperties}(b) represents the average of the effective mass extracted at different values of magnetic fields and the standard deviation determines the error bar. The backgate is progressively changed while the topgate is adjusted to keep the density constant. Using the effective mass as input, we then extract \cite{lei_electronic_2020, lodari_lightly_2022} the quantum lifetime \subtxt{\tau}{q} and the $g$-factor as shown in \autoref{fig:stateproperties}(b) and (c), respectively. We can discern a trend that $\mstar$ and \subtxt{\tau}{q} increase while the $g$-factor decreases with the magnitude of the applied backgate voltage. A power-outage interrupted the measurements before all the high-field data for the $g$-factor could be measured. 

In general, the properties of the ground state delicately depend on the confinement potential and its symmetry \cite{bir_symmetry_1974, winkler_spin_2003, terrazos_theory_2021}. Recent simulations \cite{wang_modelling_2022} have shown that the dependence of the $g$-factor on the applied electric field even depends on \gls{lh} states not confined in the \gls{qw}. Experimentally, it has been shown \cite{lodari_light_2019} that the effective mass decreases with density, i.e. with increasing confinement strength. This agrees well with our data, where $\mstar$ tends to grow when the confinement is weakened as the \gls{2dhg} is pulled away from the interface, see \autoref{fig:stateproperties}(a).

In conclusion, we successfully integrated a backgate into a Ge/SiGe heterostructure, providing an alternative to the topgate for charge accumulation and density control of the \gls{2dhg}. By combining top- and backgates, we maintained a constant density and we observed a trend towards a larger effective mass, smaller $g$-factor and increased quantum lifetime with a more negative backgate voltage. With the general nontrivial dependencies of state properties on the applied electric field, the backgate introduces an important knob for independently tuning the properties and the density. Such control will have important applications in qubit state engineering and, potentially, qubit manipulation. Furthermore the backgate facilitates precise control over bilayer \gls{qw} devices and enables compressiblity measurements. Bilayer \glspl{qw} are receiving increased attention \cite{tosato_high_2022, tidjani_vertical_2023} due to their potential for compact qubit packing, and measurements of the compressibility allow for conclusions about the density of states and the band structure \cite{eisenstein_negative_1992, henriksen_measurement_2010}. 

\section{Experimental}
\label{sec:methods}
\begin{figure}[ht!]
        \centering
        \includegraphics[width = 1.0\textwidth]{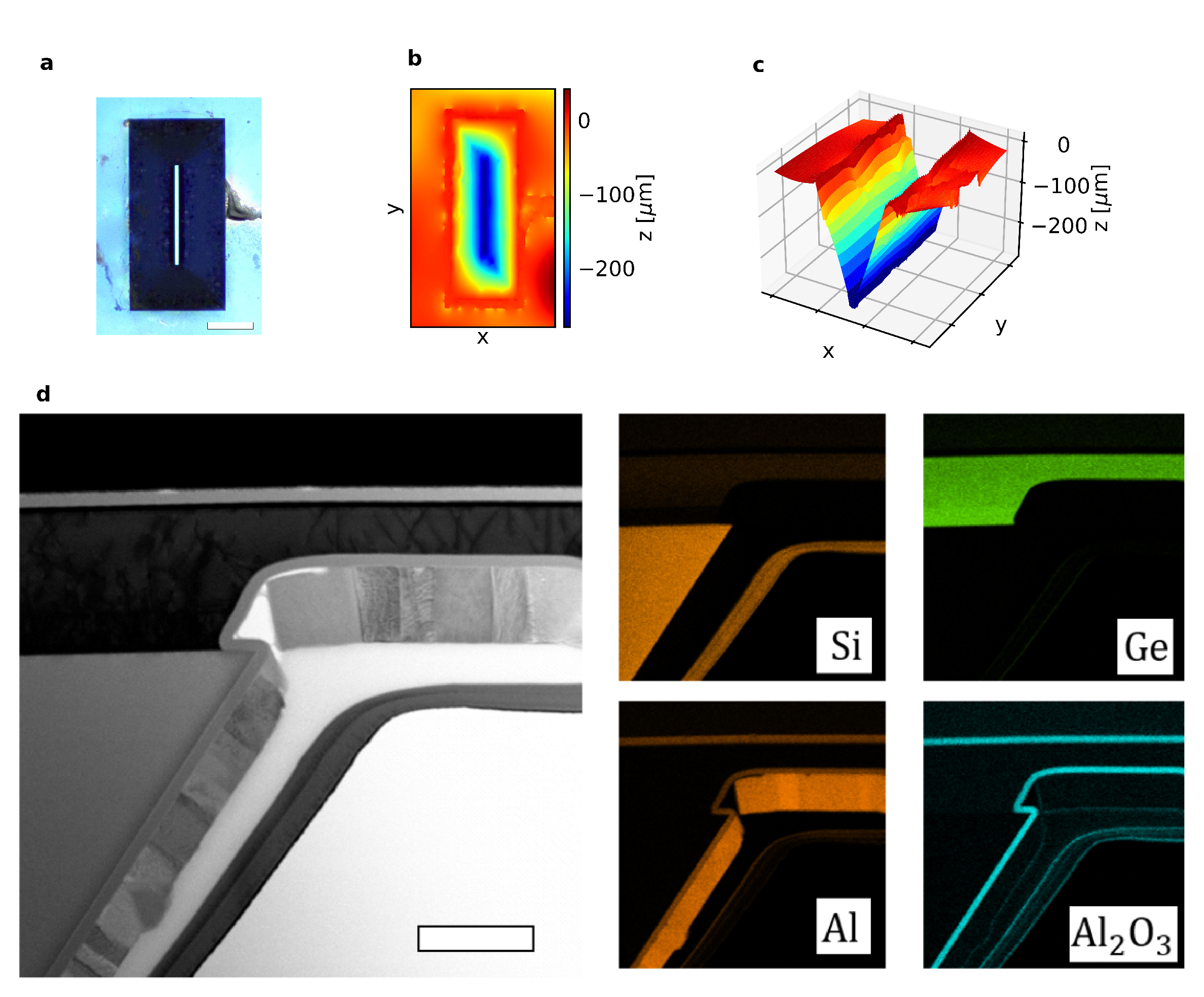}
        \caption{\textbf{Optical and \gls{tem} characterization.}(a) Confocal microscope image of the etched substrate defining the backgate region seen from the bottom. The scale bar is $\SI{200}{\micro\meter}$. (b) Height plot of (a), asymmetry along x-axis and spikes close to the edge of the etching are artifacts from the surface reflection. (c) Cut along (b) showing the typical anisotropic etching profile. (d) \textit{left} \Gls{haadf} image of a lamella cut performed perpendicular to the backgate region. The scale bar is $\SI{1}{\micro\meter}$. \textit{right} \Gls{edx} analysis of the backgate composition with the main materials involved.
        }
        \label{fig:fabrication}
\end{figure}
The backgate is fabricated by wet-etching the Si substrate with a \SI{20}{\percent} NaOH in deionized water solution at \SI{80}{\degreeCelsius} for \SI{4}{\hour}. This leads to a complete removal of the \SI{280}{\micro\meter} Si substrate and the Ge virtual substrate. During the etching process, the top surface is protected using a glass slide that is held in place with an O-Ring. The back is covered in plasma-enhanced \gls{cvd}-grown silicon nitride to which the mask is defined by standard photolithography techniques and dry etching in a \gls{rie} using a mixture of CHF$_{\mathrm{3}}$ and O$_{\mathrm{2}}$. As shown in the optical microscope and the confocal microscope images in \autoref{fig:fabrication}(a) and (b)-(c), respectively, the anisotropic etching causes the actual backgate region to be much smaller than the mask. After etching, the whole sample back is covered by aluminium oxide and a thick layer of aluminium which serves as the backgate electrode. The \gls{haadf} and \gls{edx} images show the angled etch profile and the flat bottom achieved with the NaOH etching. The etching anisotropy calls for a rectangular backgate shape and the thickness of remaining heterostructure limits the size of the backgate.

The same wet etching technique combined with \gls{rie} etching is applied to carve holes through the whole substrate, which serve as alignment marks. The achieved alignment precision is on the order of \SI{1}{\micro\meter}.

The heterostructure has been grown with \gls{cvd} as described in Refs.~\cite{nigro_high_2024, nigro_demonstration_2024}. The platinum ohmic contacts and the aluminium topgate have been fabricated using standard photolithography and lift-off techniques as outlined in Refs.~\cite{nigro_high_2024, nigro_demonstration_2024}.

\begin{acknowledgement}
The authors thank Giulio de Vito for his contributions to the NaOH etching process, Christian Olsen and Christian Schönenberger for helpful discussions and Alexander Vogel and Marcus Wyss for assistance with the \gls{tem} and \gls{edx} images.

This work has been supported by the Swiss National Science Foundation through the NCCR SPIN (grant no. 51NF40-180604).

\end{acknowledgement}

\bibliography{backgate}

\newpage
\begin{suppinfo}

\section{Supplementary S1: Hall density}
\begin{figure}[ht!]
        \centering
        \includegraphics[width = 1.0\textwidth]{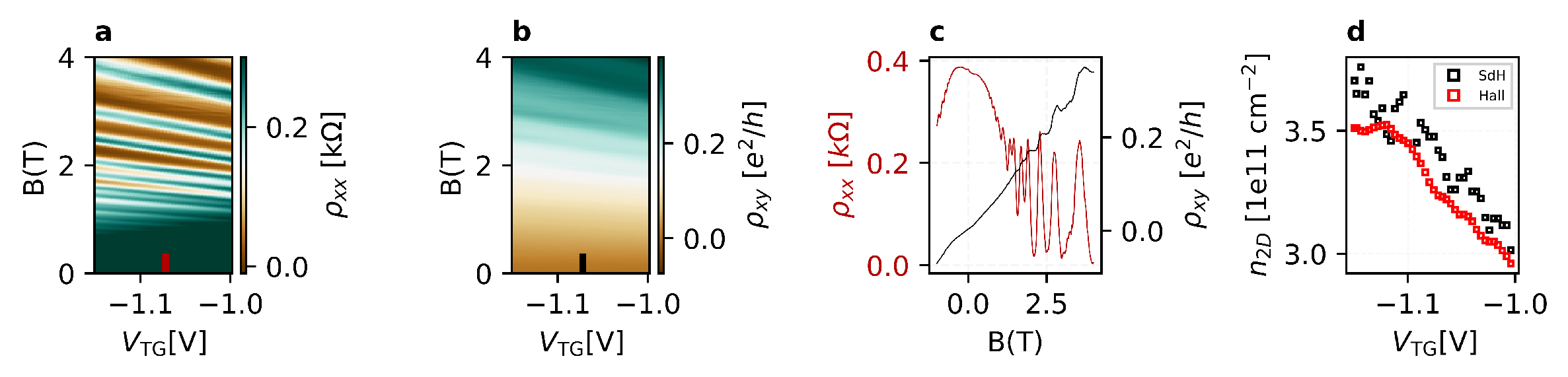}
        \caption{\textbf{Comparison of extracted Hall and \gls{sdh} density.} (a-b) hall measurements showing \subtxt{\rho}{xx} and \subtxt{\rho}{xy} as function of \vtg \,. (c) Line trace cut along the respective 2D maps for \subtxt{\rho}{xx} and \subtxt{\rho}{xy}. (d) Extracted density of holes from both \gls{sdh} oscillations and hall resistivity.} 
        \label{fig:fig_sHall}
\end{figure}

\autoref{fig:fig_sHall} shows a comparison of the hole density extracted from Hall resistivity and \gls{sdh} oscillations. \autoref{fig:fig_sHall}(a) and (b) respectively show \subtxt{\rho}{xx} and \subtxt{\rho}{xy} maps as a function of $\vtg$ and B, with voltage dependent plateaus and \gls{sdh} oscillations. The line traces in \autoref{fig:fig_sHall}(c) highlight a small mixing of longitudinal and transversal component: a slight asymmetry around zero field for \subtxt{\rho}{xx} and non-linearity in \subtxt{\rho}{xy}. Nevertheless, the extracted densities agree very well, as shown in \autoref{fig:fig_sHall}(d). The mixing could well be an artifact from the gate design which is not optimal for measuring the Hall effect. Since, with our gate design, the longitudinal resistivity is more reliable, we extract the density from \gls{sdh} oscillations throughout the whole work.


\section{Supplementary S2: \vtg \, sweep}
\begin{figure}[ht!]
        \centering
        \includegraphics[width = 1.0\textwidth]{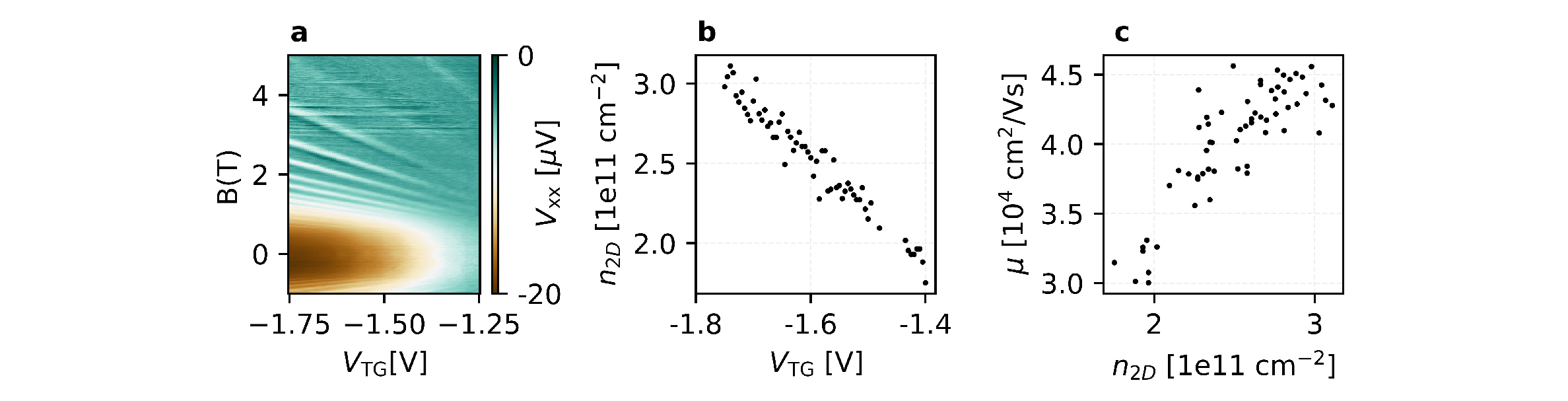}
        \caption{\textbf{Magnetotransport measurements sweeping \vtg \,.}(a) 2D map of \subtxt{V}{xx} as function of \vtg \,. (b-c) Extracted density and related mobility}
        \label{fig:fig_sVtg}
\end{figure}

We show in \autoref{fig:fig_sVtg}(a) the magnetotransport measurements obtained using only the topgate with the backgate voltage fixed at \SI{0}{\volt}. The density is obtained from the \gls{sdh} oscillations and plotted as a function of $\vtg$ in \autoref{fig:fig_sVtg}(b). The mobility, shown in \autoref{fig:fig_sVtg}(c) is evaluated using $\subtxt{\rho}{xx}(B=0)$ and the density. As discussed in the main text, the mobility versus density behaviour is independent of the gate used to measure the data.

\section{Supplementary S3: Fitting routines for $\mstar$ and \subtxt{\tau}{q}}
\begin{figure}[ht!]
        \centering
        \includegraphics[width = 1.0\textwidth]{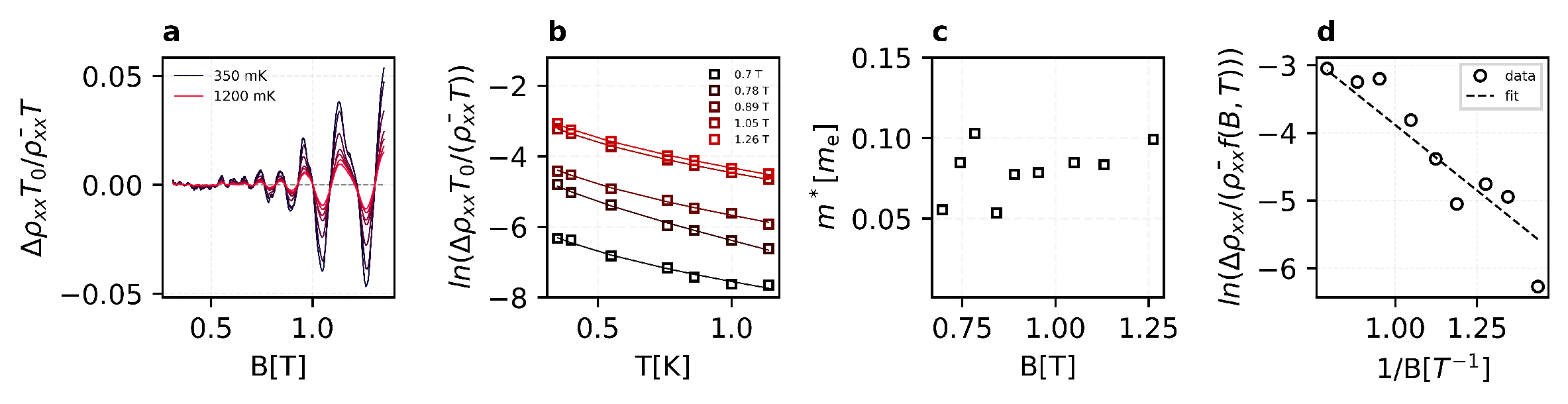}
        \caption{\textbf{\Gls{qw} properties fitting procedure.}(a) \gls{sdh} with removed magnetoresistance background for varying temperatures. (b) Fit of the envelope function for different filling factors (only selected magnetic field values shown). (c) Extracted effective mass $m^*$ for different filling factors. (d) Fit of the envelope function at lowest temperature in order to extract \subtxt{\tau}{q}.}
        \label{fig:fig_sFit}
\end{figure}

\autoref{fig:fig_sFit} displays the additional data used for fitting $\mstar$ and \subtxt{\tau}{q} as described in Ref.~\cite{lei_electronic_2020}.
\end{suppinfo}

\end{document}